\documentclass[twocolumn,showpacs,preprintnumbers,amsmath,amssymb]{revtex4}
\usepackage{graphicx}
\usepackage{epsf}
\begin{document}

\title{Collective excitations in electron-hole bilayers}

\author{G. J. Kalman$^1$, P. Hartmann$^2$, Z. Donk\'o$^2$, K. I. Golden$^3$}

\affiliation{$^1$Physics Department, Boston College,
Chestnut Hill, MA 02467, USA}

\affiliation{$^2$Research Institute for Solid State Physics and Optics of the
Hungarian Academy of Sciences,\\
H-1525 Budapest, P.O. Box 49, Hungary}

\affiliation{$^3$Department of Mathematics and Statistics and
Department of Physics, \\
University of Vermont, Burlington, VT 05401, USA}

\date{\today}

\begin{abstract}

We report a combined analytic and Molecular Dynamics analysis of the
collective mode spectrum of an electron-hole (bipolar) bilayer in the
strong coupling quasi-classical limit. A robust, isotropic energy gap
is identified in the out-of-phase spectra, generated by the combined
effect of correlations and of the excitation of the bound dipoles; the
in-phase spectra exhibit a correlation governed acoustic dispersion
for the longitudinal and transverse modes. Strong nonlinear generation
of higher harmonics of the fundamental dipole oscillation frequency
and the transfer of harmonics between different modes is observed. The
mode dispersions in the liquid state are compared with the phonon
spectrum in the crystalline solid phase, reinforcing a coherent
physical picture.

\end{abstract}

\pacs{PACS: 52.27.Gr, 52.65.Yy, 73.21.-b, 73.20.Mf }

\maketitle

Electron/hole bilayer systems consisting of a pair of two dimensional
(2D) layers of electrons and holes of equal density, (bipolar bilayers
-BPBL-s) have attracted a great deal of attention over the past decade
\cite{r01}. The main feature of the system is that in the appropriate
parameter domain, the electrons and the holes bind to each other in a
dipole-like excitonic formation \cite{r01,r02}. These excitons may
also form a Bose condensate \cite{r01,r03,r04,r05}. It is also
expected that at strong enough coupling the system undergoes a
transition into a Wigner crystal-like solid phase \cite{r02,r03}.  A
recent work has also pointed out the exciting possibility of the
existence of a supersolid phase over the dipole-solid
domain \cite{r06}. A study of a strongly coupled classical BPBL, based
on Monte Carlo simulation \cite{r02} revealed the existence
of dipole-liquid, dipole-solid, Coulomb-liquid and Coulomb-solid
``phases'', depending on the value of the coupling parameter and the
layer separation $d$. While the details of the boundaries and
transitions between the different phases would be different in the
classical and quantum domains, there is little doubt that the topology
of the classical phase diagram is quite generally correct \cite{r02}.

The problem we address in this Letter is the collective mode spectrum
of the BPBL in the strong coupling (SC) regime, in a parameter range
spanning all four phases but focusing on the more important and more
intriguing SC liquid state. The question of affinity of this spectrum
with that of the electronic bilayer (EBL), where the charges in the
two layers are identical, arises; this latter system is now fairly
well understood \cite{r07,r08,r09}. From the point of view of the
collective mode spectrum, the dominant feature of both systems in the
SC regime is the localization or quasilocalization of the particles,
either in the crystalline solid or in the SC liquid phase \cite{r10}.
This fact and the physical separation of the two oppositely charged
layers that prevents their collapse, allows one to represent the
dynamics of the collective modes in the BPBL, even in the quantum
domain, through a classical modeling \cite{r07,r10,r11}. The in-layer
exchange effects, precisely because of the quasilocalization of the
charges, contribute only to generating an exchange-correlation energy,
which is well emulated by the equivalent classical correlation energy
\cite{r12}, and tunneling is insignificant, because for any reasonable
layer separation $d$ value, $d/a_B \gg 1$. This latter condition also
ensures that the excitation energy of the exciton is well described in
terms of its classical Kepler frequency.

Our approach is based on a combined classical Molecular Dynamics (MD)
simulation and theoretical analysis. To study collective excitations,
simulation of a large assembly of particles is needed, which, at
present, becomes feasible only by using a classical approach. What
the classical approach obviously misses, however, is the effect of
condensate on the dispersion. This is further discussed in our
conclusions.

While the hole and electron masses are, generally speaking, unequal, a
simplified symmetric model with equal masses gives insight into the
dominating aspects of the mode spectrum. Within the classical
modeling, the symmetric BPBL system is then characterized by (i) the
in-layer Coulomb coupling coefficient $\Gamma = (e^2 / a k T)$ [$a = (n
  \pi)^{-1/2}$ is the Wigner-Seitz (WS) radius, $n$ is the areal
  density of particles], and (ii) the separation of the two layers
$d$.  Addressing first the theoretical description of the collective
mode spectrum, one should realize that, in contrast to the case of the
EBL, a perturbative approach is bound to fail -- even at low $\Gamma$
values -- due to the dominance of the excitonic bound state in the
spectrum. Nevertheless, as a matter of orientation, we can start with
the Random Phase Approximation (RPA) for the mode spectrum. The
interaction between the particles can be characterized by the matrix
$\varphi_{11}({\bf k}) = \varphi_{22}({\bf k}) = 2\pi e^2/k$,
$\varphi_{12}({\bf k}) = \varphi_{21}({\bf k}) = -(2\pi e^2/k)
\exp(-kd)$.  Its diagonalization in the layer space leads to the
identification of 2 longitudinal modes as in-phase and out-of-phase
($\pm$) modes: $\omega_\pm^L(k)=\omega_0\sqrt{ka}
\left[1\mp\exp(-kd)\right]^{1/2}$, where
$\omega_0=\sqrt{2e^2/ma^3}$. The small-$k$ behavior of the $+$ mode is
acoustic, here with a slope $s_+=\omega_0\sqrt{ad}$, while the $-$
mode has the typical 2D $\omega \propto \sqrt{ka}$ behavior. Thus, the
RPA misses, as it must, the appearance of the frequency of the
intrinsic oscillation (excitation) of the dipole in the spectrum. In
addition, the RPA it also unable to account for the SC collective
effect, in particular, the generation of transverse shear modes.  In
an appropriate SC description, both of these effects should be
correctly represented.  The approach we adopt is the Quasilocalized
Charge Approximation (QLCA) \cite{r10,r13}, the method that we
previously employed to identify the gapped excitation in the
collective spectrum of the EBL \cite{r07}.  The QLCA approach has also
been applied to a variety of Coulombic and other systems in the SC
liquid state \cite{r14}, with results that have been successfully
tested against the outcomes of computer simulations \cite{r08,r09} and
laboratory experiments \cite{r12,r15,r16} in both the classical and
quantum domains. The longitudinal ($L$) and transverse ($T$) QLCA
dielectric matrices in layer space ($i,j$) \cite{r07}
\begin{equation}
\varepsilon_{ij}^{L,T}({\bf
  k},\omega)=\delta_{ij}-\frac{nk^2}{m}\varphi_{il}({\bf
  k})\cdot\left[\omega^2{\bf I} - {\bf D}^{L,T}({\bf
    k})\right]_{lj}^{-1},
\end{equation}
become functionals of the longitudinal ($L$) and transverse ($T$)
projections of the dynamical matrix $D_{ij}^{\mu\nu}({\bf
  k})=\frac{1}{mA}\sum_{\bf q}q^\mu q^\nu
\varphi_{ij}(q)\left[h_{ij}\left(|{\bf k} - {\bf
    q}|\right)-\delta_{ij}\sum_l\varphi_{il}(q)h_{il}({\bf q})\right]$,
  $h_{ij}({\bf q})$ are the Fourier transforms of
    $h_{ij}(r)=g_{ij}(r)-1$, where $g_{ij}(r)$ is the pair
    distribution function and $A$ is a large but finite surface. The
    resulting mode structure
\begin{eqnarray}
\omega_\pm^L({\bf k})=\left[\omega_0^2 ka \left(1\mp\exp(-kd)\right) +
  D_{11}^L({\bf k}) \pm D_{12}^L({\bf k})\right]^{1/2}
\\ \omega_\pm^T({\bf k})=\left[D_{11}^T({\bf k}) \pm D_{12}^T({\bf
    k})\right]^{1/2}
\end{eqnarray}
now exhibits a behavior significantly different from its RPA
counterpart. These dispersion curves are displayed in Fig.~\ref{fig1}
together with MD simulation results (see below).

\begin{figure}[ht]
\includegraphics[scale=0.6]{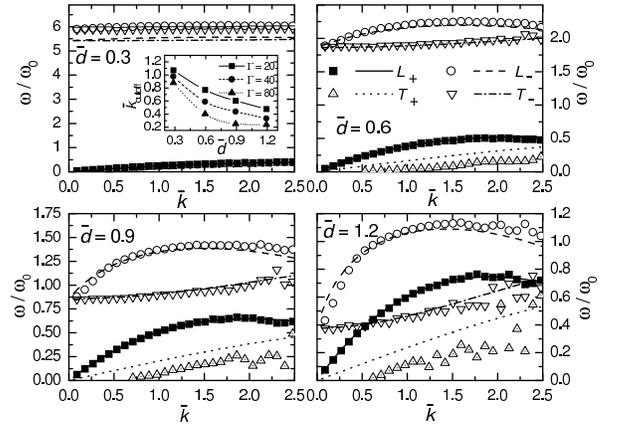}
\caption{Dispersion relations for the $L_+$, $L_-$, $T_+$, and $T_-$
  modes ($\bar{k}=ka$) at $\Gamma = 40$ and different layer
  separations. The legend shown in the $\bar{d}=d/a=0.6$ plot applies
  to all panels. Lines: QLCA, symbols: MD simulation. Inset shows the
  observed values of $\bar{k}_{\rm cutoff}$ of the $T_+$ mode.
\label{fig1}}
\end{figure}

\begin{figure}[ht]
\includegraphics[scale=0.65]{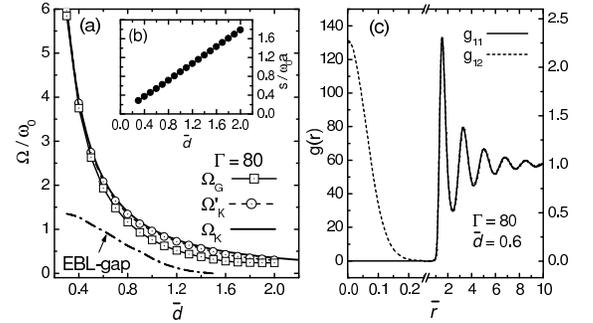}
\caption{(a) Dependence of the Kepler frequency $\Omega_K$, the
  thermally shifted Kepler frequency $\Omega^\prime_K$ and the gap
  frequency $\Omega_G$ on $\bar{d}$ at $\Gamma = 80$; also shown for
  comparison is the EBL gap. (b) $s_+$ sound velocity vs. $\bar{d}$
  for $\Gamma = 80$. (c) MD pair correlation functions for $\Gamma =
  80$ and $\bar{d} = 0.6$ (left and right vertical scales belong to
  the left and right horizontal parts).
\label{fig2}}
\end{figure}

(i) At $k=0$ the $-$ mode becomes a gapped mode (see also
Fig.~\ref{fig2}a), with
\begin{equation}
\omega_-^L(k=0)=\omega_-^T(k=0) \equiv \Omega_G(d)
\end{equation}
where $\Omega_G(d)$ is a functional of $h_{12}(r)$ as given in
\cite{r17}. In the BPBL $h_{12}(r)$ is governed by a central peak
$h_{12}^\prime(r)$ around $r=0$ \cite{r02} (see Fig.~\ref{fig2}c). For
$d/a<1$ $\Omega_G(d)$ is well described by replacing $h_{12}(r)$ by
$h_{12}^\prime(r)$. When this central peak is approximated by a
Gaussian (representing a thermally excited dipole)
$\Omega_G(d)$ becomes $\Omega_K^\prime(d)$, the thermally broadened small
amplitude oscillation Kepler frequency $\Omega_K(d)=\sqrt{2e^2/d^3m}=
\omega_0(a/d)^{3/2}$ of the oscillating dipole. Dipole-dipole
correlations shift $\Omega_G(d)$ from this value only for $d/a>1$.

(ii) The $+$ mode may be regarded as a density oscillation of the
dipoles: with a $1/r^3$ type dipole-dipole interaction, this suggests
an $\omega \propto k$ dispersion. This is borne out by the present
calculation: the $+$ mode shows an acoustic behavior, similarly to the
RPA result, but with a phase velocity instead of being proportional to
$(d/a)^{1/2}$, is of the order of $d/a$:
$s\equiv\omega_+^L(k\rightarrow 0)/k=\omega_0 d \left[(99/96)\int {\rm
    d}\bar r g_{12}(r)/\bar r^2 \right]^{1/2}$ (see Fig.~\ref{fig2}b).

(iii) In addition to the 2 longitudinal modes 2 transverse shear modes
appear, with a behavior qualitatively similar to their longitudinal
counterparts, except that, as expected for shear modes in a liquid,
the $T_+$ mode do not extend below a finite $k_{\rm cutoff}$ value
(see inset in Fig.~\ref{fig1}) \cite{kcut}.

(iv) For $k\rightarrow \infty$ all four modes approach the Einstein
frequency (the frequency of oscillation of a single particle in the
frozen environment of the others) \cite{r07} of the system:
\begin{equation}
\Omega_{E}^2(d)=\Omega_{E,2D}^2+\frac{1}{2}\Omega_{G}^2(d),
\label{eq_6}
\end{equation}
where $\Omega_{E,2D}^2=(\omega_0^2/2)\int{\rm
d}\bar r  g_{11}(r)/\bar r^2 \cong 0.38 \omega_0^2$ is the Einstein
frequency of an isolated 2D layer \cite{r14} (see later).

\begin{figure}[ht]
\includegraphics[scale=0.65]{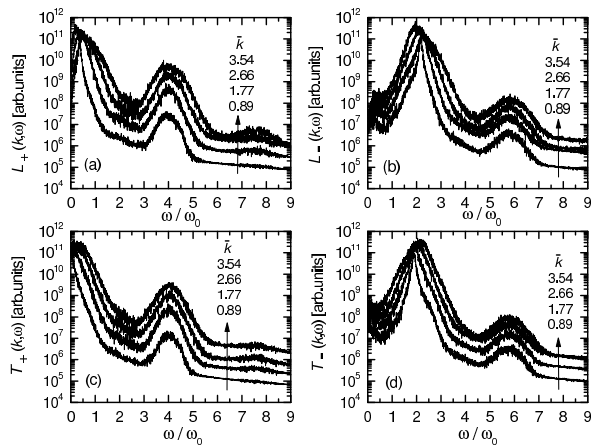}
\caption{Longitudinal current (a,b) and transverse current (c,d)
  fluctuation spectra obtained at $\Gamma = 40$ and $\bar{d} =
  0.6$. The arrows indicate increasing wave numbers $\bar{k}$ as
  listed in the panels. Note the appearance of higher harmonics of the
  gap frequency: the second and fourth harmonic in the $+$ spectra,
  and the third harmonic in the $-$ spectra.
\label{fig3}}
\end{figure}

Illustrative longitudinal and transverse current-current correlation
spectra obtained from our MD simulations (based on the
Particle-Particle Particle-Mesh (PPPM) method \cite{r18} with periodic
boundary conditions) are displayed in Fig.~\ref{fig3} for $\Gamma =
40$ and $d/a = 0.6$. Collective modes show up as peaks in the
correlation spectra. The MD-generated dispersion relations inferred
from the spectral peaks for the 4 modes for $\Gamma = 40$ and for a
series of $d/a$ values are displayed in Fig.~\ref{fig1}, together with
theoretically calculated values. Agreement between the QLCA theory and
simulation data is very good, especially for lower $k$
values. Deviation from the characteristic $d^{3/2}$ behavior of
$\Omega_K(d)$ seems to occur on passing the (dipole liquid)/(Coulomb
liquid) phase boundary \cite{r02}. Comparison with the EBL spectrum
\cite{r07,r08} reveals a more pronounced and more robust gap value
(Fig.~\ref{fig2}a), which is not unexpected if one is mindful of the
different physical mechanisms responsible for creating the gap.

\begin{figure}[ht]
\includegraphics[scale=0.7]{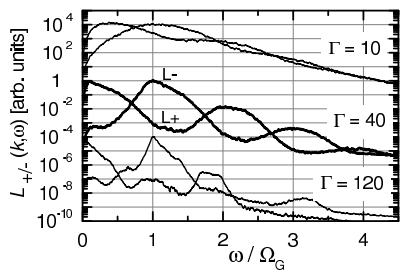}
\caption{Longitudinal current fluctuation spectra showing the higher
  harmonics of the gap frequency for $\bar{k}=3.54$, $\bar{d} = 0.6$,
  $\Gamma=10$, 40 and 120. The transverse spectra exhibit similar
  behavior.
\label{fig4}}
\end{figure}

A remarkable feature of the simulation spectra is the appearance of
higher harmonics of the gap frequency: we observe the emergence (a) of
the third harmonic in the $-$ spectra and (b) of the second harmonic
({\it but not the fundamental}) in the $+$ spectra)
[Fig.~\ref{fig4}]. These harmonics are the most pronounced around
$\Gamma = 40$ in the strongly coupled dipole liquid phase, showing a
diminishing trend both for lower and higher coupling values: at high
$\Gamma$ values, the low amplitude of the thermally excited
oscillations is not conducive to the generation of harmonics, while at
lower $\Gamma$ values, the thermal motions damp the higher
harmonics. The analysis of the details of the nonlinear processes
underlying the harmonic generation will be presented elsewhere. 

We calculate the phonon spectrum of the crystalline phase, through the
standard harmonic approximation, by summing over a lattice of $2 \times
10^7$ sites. In the solid phase the BPBL crystallizes in a hexagonal
structure, with the particles in the two layers facing each other (for
any layer separation, in contrast to the EBL which exhibits a variety
of lattice structures).  One can recognize the 4 modes identified in
the liquid state, with an expected anisotropic dispersion and with the
understanding that the ``longitudinal'' and ``transverse'' labeling of
the modes describes their polarization for propagation along the
principal crystal axes only.

\begin{figure}[ht]
\includegraphics[scale=0.75]{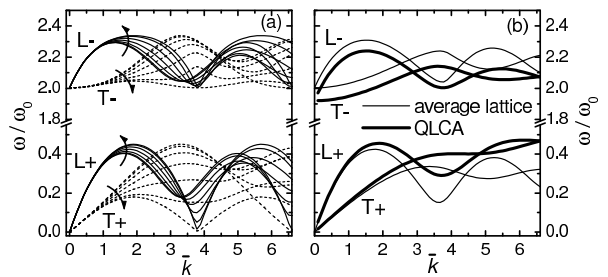}
\caption{(a) Lattice dispersion of the quasi-$L_+$ and quasi-$L_-$
  (full lines) and quasi-$T_+$ and quasi-$T_-$ (dashed lines) modes
  for a series of lattice angles between $0^o \leq \alpha \leq 30^o$
  ($\alpha=0^0$ corresponds to the nearest neighbor direction, the
  arrows indicate increasing $\alpha$). (b) Comparison at
  $\bar{d}=0.6$ of QLCA (at $\Gamma=80$) and angular averaged lattice
  dispersion.
\label{fig5}}
\end{figure}

The mode spectrum is shown in Fig.~\ref{fig5}. For the sake of
comparison with the QLCA dispersion, also shown are the angle-averaged
longitudinal and transverse projections: the similarity between the
two is the consequence of the underlying microscopic order in the SC
liquid as predicted by the QLCA. The $-$ modes in the lattice spectrum
exhibit a $k = 0$ gap, very similar to the one in the liquid spectrum;
in contrast to the EBL, the gap is not split by the lattice
anisotropy.  This is a consequence of the quasi-isotropy of the
hexagonal lattice environment to $O(k^2)$. Physically, the $-$ modes
can be regarded, in a good approximation, especially for small $d/a$
values, as the superposition of a $+$ mode on the $\Omega_K$ dipole
oscillation frequency. The angle-averaged modes emulate the
liquid-phase mode dispersion for low and moderate $k$-values, but
deviate substantially from it for $k \rightarrow \infty$ (see
Fig.~\ref{fig5}b). To understand this difference, one has to focus on
$h_{12}^\prime(r)$, which shows a quasi-Gaussian behavior in the
liquid, but becomes a delta-function in a perfect $T = 0$ lattice. The
contribution of oscillating dipoles within a Gaussian distribution
results in a superposition of random phases at $k \rightarrow \infty$,
averaging out to 0, ensuring the limit shown in Eq.~\ref{eq_6}. In
contrast, the delta-function contribution generates additional
coherent $k$-independent terms that survive for $k \rightarrow
\infty$.  The simulation results clearly indicate that even for high
$\Gamma$ values the large but finite $k$ dispersion curves follow the
trend set by Eq.~\ref{eq_6} describing the liquid, rather than the one
corresponding to the lattice. The great sensitivity of the behavior of
the mode dispersions at higher $k$ values to the width of
$h_{12}^\prime(r)$ could motivate an observational technique for
acquiring information of the width of central peak distribution
through tracking the high-$k$ behavior of the collective excitations.

The phonon spectrum in the BPBL was the subject of a recent
investigation by Ref. \cite{r20}. This work addressed the behavior of
the dipolar ``phase'', within an approximation geared to the assumed
separability of the overall collective and intrinsic dipolar
excitations: our results show that this technique, reasonable as it
may appear, leads to an unphysical behavior in the calculated mode
dispersion.  In particular, \cite{r20} finds (i) that the gap is split
and there are distinct transverse and longitudinal gap values; (ii)
that both of these exhibit an anisotropic behavior; (iii) that for $k
\rightarrow 0$ the $-$ mode dispersion curves exhibit a negative
slope; and (iv) that the $k \rightarrow \infty$ behavior does not
conform to that demanded either by Eq.~\ref{eq_6} or by the lattice
condition.  These features violate basic physical principles pointed
out above.

In summary, we have determined the dispersion characteristics of a
strongly coupled symmetric bipolar (electron/hole) bilayer in the
strongly coupled domain, where the dominant exchange-correlation
energy provides a sufficient quasilocalization of the particles to
engender a quasi-classical behavior. Our analysis in the strongly
coupled liquid phase is based on the MD simulation of the density and
current fluctuation spectra and on the application of the theoretical
QLCA technique. The benchmark phonon spectrum in the crystalline solid
phase is determined through lattice summation technique. The results
of all these various approaches reinforce each other and provide a
coherent physical picture. The 4 modes, characteristic of bilayer
systems (in-phase/out-of-phase, longitudinal/transverse) emerge, with
the expected $k = 0$ energy gap in the spectrum of the out-of-phase
mode \cite{r07,r08}. The latter, is now mostly governed by the
intrinsic oscillation frequency of the dipoles and affected by
collective interaction only in the Coulomb phase, for $d \gtrsim a$. A
remarkable effect, unique to the strongly coupled liquid phase, the
generation of harmonics of the gap frequency and the transfer of the
even harmonics from the out-of-phase mode to the in-phase mode, has
been observed.

The presence of the condensate would manifest itself primarily by
reducing the quasi-localization of the particles (as clearly
demonstrated by \cite{r21}). A rough estimate of the degree of this
reduction is provided by replacing $h(k)$ (and thus $D(k)$) by
$\bar{h}(k)=(1-f)^2h(k)$, where $f$ is the condensate fraction, as
suggested by \cite{r22}. Since $f$ in the SC regime is expected to be
quite small ($f=0.02$ \cite{r05}), this should not be a dramatic
effect. In the solid phase, if the supersolid develops, the system
would probably behave as a binary mixture of the normal crystal
lattice and the superfluid liquid constituted by the vacancies
\cite{r23}, with a coupling between the lattice phonons and the
collective excitations of the latter. These issues have to be
addressed elsewhere.

This work was supported by NSF Grants PHY-0514619 (GJK), PHY-0514618
(KIG), and by the Hungarian Fund for Scientific Research and the
Hungarian Academy of Sciences, OTKA-T-48389, MTA-OTKA-90/46140,
OTKA-PD-049991.

\end{document}